\documentclass[amsmath, amssymb, aps, pra,twocolumn,floatfix,nofootinbib]{revtex4}
\usepackage[utf8]{inputenc}
\usepackage{titlesec}
\usepackage[colorlinks]{hyperref}
\usepackage{graphicx}
\usepackage{amsmath,amssymb}
\usepackage{bbold}
\usepackage{float}
\usepackage{dcolumn}
\usepackage{natbib}
\usepackage{bm}
\usepackage{dsfont}
\usepackage{url}
\usepackage{mathtools}
\usepackage{amsthm}
\usepackage{amsfonts}
\usepackage{braket}
\usepackage{cancel}
\usepackage{bbold}
\usepackage{makecell}
\usepackage[normalem]{ulem}
\usepackage{xcolor}
\usepackage{comment}

\usepackage{afterpage}

\begin{document}
\title{Ensemble basis-free coherence for quantum communication}
\author{A.\,Kodukhov$^{*}$}
\affiliation{\text{Terra Quantum AG, Kornhausstrasse 25, St. Gallen, Switzerland, 9000}
\\$^*\,$ak@terraquantum.swiss
}
\begin{abstract}
The resource theory of coherence quantifies the quantum properties of states and their role in quantum information processing.
Extending coherence theory to ensembles, the natural objects in quantum communication, remains an active research direction, with their operational potential still being explored.
In this work, we establish a maximally coherent ensemble in every finite dimension: the continuous ensemble of pure states distributed according to the normalized Haar measure attains the tight upper bound on basis-free relative entropy ensemble coherence.
We then prove that ensemble coherence bounds the effective informational advantage of postselective measurements over the Holevo quantity for two equiprobable pure states.
Finally, for the qubit B92 QKD protocol, we show that the coherence of Alice's ensemble upper-bounds the secret key rate in the asymptotic regime.
These results give ensemble coherence a quantitative operational meaning in quantum communication and cryptography.
\end{abstract}

\maketitle
\section{Introduction}
Quantum coherence turns the superposition principle into a resource for quantum information processing\,\cite{streltsov2017colloquium}.
In the framework of Baumgratz et\,al.\,\cite{baumgratz2014quantifying}, the relative entropy of coherence quantifies this resource with respect to a chosen basis, while incoherent operations specify the transformations that cannot generate it from incoherent states.
When information is encoded into an ensemble of quantum states, coherence can also be studied as a collective property of the ensemble, independent of the choice of reference basis.
Studies of ensemble quantumness address this question by minimizing the average relative entropy of coherence over a common basis\,\cite{luo2010quantum,kronberg2019coherence,mao2019quantumness}.
The entanglement-based approach gives this basis-free quantity a physical interpretation, connecting the coherence of ensembles generated by local measurements to entanglement and measurement uncertainty\,\cite{kodukhov2023entanglement,kodukhov2025converting}.
This perspective raises two fundamental questions: what are the ensemble counterparts of maximally coherent states and incoherent operations, and how does ensemble coherence constrain the information that measurements can extract?

Quantum communication gives these questions immediate operational relevance: classical bits are encoded into ensembles, and their quantum properties determine how much information a receiver or an eavesdropper can extract.
In particular, quantum key distribution (QKD) calls for identifying which quantum resources are necessary or sufficient for secrecy.
Entanglement distillation provides a route to secure key generation\,\cite{lo1999unconditional}, but the relevant resources depend on the cryptographic setting.
Pirandola\,\cite{pirandola2014quantum} identified quantum discord as a necessary resource and showed that device-dependent QKD can remain secure even without shared entanglement.
Ref.\,\cite{wang2026witnessing} developed this connection further by using an experimentally accessible quantum discord witness to lower-bound the secret key rate of the three-state QKD with trusted but imperfect devices.
A complementary connection was established by Ma et\,al.\,\cite{ma2019operational}, who related secret key rates under collective attacks to the coherence of shared bipartite states.
These results motivate exploring the quantitative meaning of basis-free ensemble coherence in quantum cryptography.

In this work, we investigate the fundamental properties of ensemble coherence and its operational role in quantum communication.
In Section\,\ref{sec:basis}, building on Haar-averaged coherence\,\cite{singh2016average,cheng2015complementarity,zhang2017average}, we establish a tight upper bound on the ensemble coherence that depends only on the Hilbert space dimension.
We show that the continuous ensemble of all pure states distributed according to the normalized Haar measure attains this bound, identifying a maximally coherent ensemble in every finite dimension.
Following the characterization of commutativity-preserving channels\,\cite{yu2011quantum,streltsov2011behavior,hu2012necessary}, we define these channels as the nonselective incoherent operations for ensembles.
The monotonicity of relative entropy ensemble coherence under these operations was established in Ref.\,\cite{mao2019quantumness}.

In Section\,\ref{sec:intercet-resend}, two examples arising from quantum communications, a measure-and-prepare transformation and a beam splitter-like transformation\,\cite{kronberg2025local}, clarify how this notion differs from incoherent operations in the Baumgratz framework.
These examples illustrate that the operational value of coherence generation depends on the communication setting: coherence created during trusted preparation can support subsequent information processing, whereas its increase during transmission can accompany information leakage, reduced signal distinguishability, or loss of entanglement between Alice and Bob.

We then establish the operational role of ensemble coherence in the information extraction and secret key generation in Section\,\ref{sec:coherence_bounds}.
For coherent ensembles of pure states, postselection enables a conditional informational advantage over the Holevo quantity.
Building on Ref.\,\cite{kenbaev2022quantum}, we show that ensemble coherence bounds this informational advantage weighted by the probability of a conclusive measurement result for two equiprobable pure states in any finite dimension.
For the qubit B92 QKD protocol\,\cite{bennett1992quantum}, we further derive an upper bound on the secret key rate in terms of Alice's initial ensemble coherence.
These results give ensemble coherence a quantitative meaning in quantum communication and motivate further exploration of its quantum cryptographic role.
Finally, we conclude in Section\,\ref{sec:conclusion}.

\bigskip
\section{Maximally coherent ensemble and incoherent operations}\label{sec:basis}
In this section, we recall the definition of the ensemble basis-free coherence and establish its resource theoretical properties: provide maximally coherent ensemble, define incoherent operations and show monotonicity of ensemble coherence under incoherent operations.

For a single quantum state, the resource theory of coherence is based on the quantum relative entropy\,\cite{vedral2002role} and is described by the Baumgratz framework\,\cite{baumgratz2014quantifying}.
The relative entropy of coherence for a quantum state $\rho $ in a fixed orthonormal basis $T= \lbrace \ket{e_k} \rbrace_{k=0}^{d-1}$ of the dimension $d$ is defined as
\begin{equation}
    C_{T}^\text{rel}(\rho)
    =D\!\left(\rho\middle\|\rho^\text{diag}\right)
    =S\left( \rho^\text{diag}\right) - S(\rho),
\end{equation}
where $D\!\left(\rho\middle\|\rho^\text{diag}\right) = \text{tr}\left[ \rho \left( \log_2 \rho - \log_2 \rho^\text{diag} \right) \right]$ is the quantum relative entropy and $\rho^\text{diag} = \sum_{k=0}^{d-1} \bra{e_k}\rho \ket{e_k} \ket{e_k}\!\bra{e_k}$.

The basis-free coherence of an ensemble $\mathcal{E}=\{p_i,\rho_i\}$ is defined as minimized over all bases averaged over the ensemble relative entropy of coherence\,\cite{luo2010quantum,kronberg2019coherence, kodukhov2023entanglement, mao2019quantumness} 
\begin{equation}
    C(\mathcal{E})
    =\min_{T}\sum_i p_i C_{T}^\text{rel}(\rho_i).
    \label{eq:ensemble-relative-entropy}
\end{equation}
We also allow continuous ensembles.
Let $\mathcal{S}(\mathcal{H}^d)$ denote the set of density operators on a $d$-dimensional Hilbert space $\mathcal{H}^d$.
A generalized ensemble is specified by a Borel probability measure $\mu$ on $\mathcal{S}(\mathcal{H}^d)$\,\cite{holevo2004continuous}.
Its basis-free coherence is defined as
\begin{equation}
    C(\mu)=\min_T
    \int_{\mathcal{S}(\mathcal{H}^d)}
    C_T^{\mathrm{rel}}(\rho)\,\mu(d\rho),
    \,
    \int_{\mathcal{S}(\mathcal{H}^d)}\mu(d\rho)=1.
\end{equation}
For the discrete measure $\mu=\sum_i p_i\delta_{\rho_i}$, where $\delta_{\rho_i}$ is concentrated at $\rho_i$, this definition reduces to Eq.\,(\ref{eq:ensemble-relative-entropy}).

In the Baumgratz framework of coherence, the set of incoherent states with respect to the reference basis $T$ is defined as $\mathcal{I}_T = \lbrace \rho = \sum_k \bra{e_k}\rho \ket{e_k} \ket{e_k}\!\bra{e_k} \rbrace$, i.e. all the states with the eigenbasis $T$.

An ensemble $\mathcal{E} = \{p_i,\rho_i\}$ is called incoherent, if its coherence measure defined by Eq.\,(\ref{eq:ensemble-relative-entropy}) is zero, i.e. all the density matrices of the ensemble has a common eigenbasis
\begin{equation}
    \exists \, T=\lbrace \ket{e_k} \rbrace_{k=0}^{d-1} :\, \forall i \,\rho_i = \sum_{k=0}^{d-1} \bra{e_k}\rho_i \ket{e_k}\,\ket{e_k}\!\bra{e_k}.
    \label{eq:incoherent_ensemble}
\end{equation}
Alternatively, for an incoherent ensemble, all its states mutually commute
\begin{equation}
    \forall i,j\,\left[ \rho_i, \rho_j \right]=0.
\end{equation}
In contrast to the coherence framework for a single state, the definitions of the ensemble coherence  and an incoherent ensemble do not depend on the basis choice.

In the Baumgratz framework, maximally coherent state with respect to a basis $T$ is the uniform superposition of all reference-basis states $\frac{1}{\sqrt{d}}\sum_{k=0}^{d-1}\ket{e_k}$.

Following the framework presented in Ref.\,\cite{singh2016average}, we establish an upper bound on the ensemble coherence and determine maximally coherent ensemble for a fixed dimension $d$.
Consider an ensemble of pure states $\mathcal{E}=\{p_i,\ket{\psi_i}\}$ and an orthonormal basis $T = \lbrace \ket{e_k} \rbrace_{k=0}^{d-1}$.
The ensemble coherence in this basis is 
\begin{equation}
    C_T\left( \mathcal{E} \right)=\sum_i p_i H\left( \lbrace |\braket{e_k|\psi_i}|^2 \rbrace_k \right),
\end{equation}
where $H(\cdot)$ is the Shannon entropy.
Its minimum cannot exceed its average over all basis choices.
We denote the basis choice as a unitary transformation $U$ that maps the basis $T$ to another basis $T_U = UT$.
The ensemble coherence averaged over uniformly distributed bases 
\begin{multline}
    \bar{C}(\mathcal{E}) = \int_{U(d)} C_{T_U} (\mathcal{E}) dU \\
    =\sum_i p_i \int_{U(d)}   C_T^\text{rel} \left( U \ket{\psi_i} \right) dU,
\end{multline}
where $d U$ is the normalized Haar measure on the set of all unitary transformations over the Hilbert space of dimension $d$, $U(d)$.
For the qubit case, the Haar measure is reduced to $dU = \sin \theta d\theta d\phi/(4 \pi)$ for $\theta \in [ 0; \pi ]$ and $\phi \in [0;2\pi]$.

The averaged over all bases relative entropy of coherence for a pure state does not depend on the input state.
Here, we invoke the already obtained results for the averaged relative entropy of coherence\,\cite{cheng2015complementarity, luo2019average, zhang2017average}.
Since we are interested in the upper bound on the ensemble coherence, we consider only pure states case
\begin{equation}
    \forall i \int_{U}   C_T^\text{rel} \left( U \ket{\psi_i} \right) dU = \frac{L_d}{\ln 2},\,L_d = \frac{1}{2}+\frac{1}{3}+\dots+\frac{1}{d}.
\end{equation}
As a result, we obtain the upper bound for ensemble coherence minimized over all bases
\begin{equation}
    C(\mathcal{E}) \leq \bar{C}(\mathcal{E}) = \frac{L_d}{\ln 2}.
    \label{eq:coherence-upper-bound}
\end{equation}
The same bound applies to ensembles of mixed states by convexity of the relative entropy of coherence: refining every mixed state into pure states cannot decrease the corresponding upper bound.

The bound in Eq.\,(\ref{eq:coherence-upper-bound}) is attained by the continuous ensemble of all pure states equipped with the normalized Haar measure.
This ensemble is rotationally invariant, so coherence has the same value for every basis and the minimization cannot lower it.
Consequently, the Haar-uniform ensemble has $C(\mathcal{E})= L_d/\ln2$ and is maximally coherent among ensembles of the dimension $d$.
For discrete ensembles, the upper bound of Eq.\,(\ref{eq:coherence-upper-bound}) is a supremum.

Importantly, the ensemble coherence is upper-bounded by the Holevo quantity\,\cite{holevo} of an ensemble\,\cite{kodukhov2023entanglement} and for the $d$-dimension case the maximum possible Holevo value is $\log_2 d $.
As we observe, provided in Eq.\,(\ref{eq:coherence-upper-bound}) upper bound can be tighter than the Holevo bound.
The resulting combined upper bound is
\begin{equation}
    C(\mathcal{E}) \leq \min\left\lbrace \chi\left( \mathcal{E} \right), \frac{L_d}{\ln 2} \right\rbrace.
\end{equation}
Thus, entanglement cannot always be fully converted into ensemble coherence\,\cite{kodukhov2025converting}, because the initial entanglement may exceed the maximum ensemble coherence allowed by the output dimension.
For the qubit case, the maximally coherent ensemble has coherence $1/(2\ln 2)\approx0.72$ that is smaller than $1$ provided by the Holevo quantity upper bound.
In the qubit case, the maximally coherent ensemble is a continuous set of homogeneously distributed pure states covering the whole surface of the Bloch sphere.

Resource theory requires the definition of basic physical operations.
In the Baumgratz framework, nonselective incoherent operations keep the set of incoherent states invariant: $\forall \rho \in \mathcal{I}_T\,\, K_n \rho K_n^\dag/\text{tr}(K_n \rho K_n^\dag) \in \mathcal{I}_T$, where $K_n$ are the Kraus operators of the channel, satisfying $\sum_n K_n^\dag K_n = \mathds{1}$.
A quantum channel is incoherent in the Baumgratz framework if it admits a Kraus representation in which every branch maps incoherent states to incoherent states.
The nonselective output is then also incoherent, since it is a mixture of incoherent branch states.

The definition of nonselective incoherent operations for ensemble coherence is more subtle.
One can come up with the similar definition as for single density matrices: if each Kraus operator of a quantum channel does not produce ensemble coherence for a classical ensemble defined in Eq.\,(\ref{eq:incoherent_ensemble}), the channel can be considered incoherent.
However, as we show below in Example 1, the channel as a combination of all Kraus effects can still create ensemble coherence despite each Kraus operators make the output ensembles classical.
Instead, we call an operation incoherent, if for any classical input ensemble the output ensemble is also classical.
Classicality of an ensemble means that all its density matrices mutually commute.
Thus, for the ensemble coherence, all incoherent operations are commutativity preserving operations.

Importantly, the basis free ensemble coherence defined in Eq.\,(\ref{eq:ensemble-relative-entropy}) is monotone under commutativity-preserving operations\,\cite{mao2019quantumness}, i.e. ensemble coherence based on the relative entropy has monotonicity under incoherent operations as one would expect from a coherence measure.
For other ensemble quantumness measures, the monotonicity under commutativity-preserving operations was also demonstrated, in particular, for the quantumness based on the mutual commutators\,\cite{li2019monotonicity}.

Let $\Phi$ be a commutativity-preserving completely positive trace-preserving channel, that is,
\begin{equation}
    [\rho,\sigma]=0
    \quad\Longrightarrow\quad
    [\Phi(\rho),\Phi(\sigma)]=0
\end{equation}
for all density operators $\rho$ and $\sigma$. Then, for every ensemble $\mathcal{E}$,
\begin{equation}
    C\!\left(\Phi(\mathcal{E})\right)\leq C(\mathcal{E}),
    \qquad
    \Phi(\mathcal{E})=\{p_i,\Phi(\rho_i)\}.
    \label{eq:ensemble-coherence-monotonicity}
\end{equation}

The preservation of classical ensembles under commutativity preserving operations is fully equivalent to the noncreation of quantum correlations studied in Refs.\,\cite{yu2011quantum, streltsov2011behavior, hu2012necessary}.
For the relative entropy ensemble coherence, the monotonicity stated above provides the stronger quantitative guarantee that these channels cannot increase coherence even when the input ensemble is already coherent.

Incoherent operations defined above and incoherent operations in the Baumgratz framework are different sets of operations.
For example, the Hadamard gate that maps computational basis $\ket{0},\ket{1}$ to the Hadamard states $\ket{+}, \ket{-}$ create coherence for a single quantum state, while, for any input ensemble, this gate keeps ensemble coherence unchanged.
In fact, any unitary operation is incoherent for the ensemble coherence.
In the next section, we study several examples of quantum channels, their difference for ensemble coherence and Baumgratz framework, how these channels affect ensemble coherence, and study information leakage to the environment in these examples.

\bigskip
\section{Coherence and quantum channels}
In this section, we study abilities of different quantum channels to create ensemble basis-free coherence.
Considered quantum channels correspond to common transformations in quantum communication scenarios: measure-and-prepare transformation and beam-splitter like transformation.
Initial amount of coherence represent general robustness of ensembles to different type of transformations and how well ensembles keep the quantum properties associated with the superposition and indistinguishability.

A recent study\,\cite{swain2026persistence} considers triality relations among interference visibility, path predictability, and entanglement under noisy quantum channels.
Here, we investigate how channels affect basis-free ensemble coherence and information leakage to the environment.

Consider a general quantum communication scenario, in which legitimate sender, Alice, locally encodes classical information into ensemble of quantum states and, then, transmits the ensemble to a legitimate receiver, Bob. 
The classical register is denoted by A, while the subsystem of the quantum ensemble is denoted by B.
Shared Alice-Bob entangled state reads
\begin{equation}
    \ket{\Psi}_\text{AB} = \sum_{k=0}^{N-1} \sqrt{p_k} \ket{k}_\text{A} \otimes \ket{\psi_k}_\text{B},
    \label{eq:ent_state}
\end{equation}
where $N$ is the number of states in the ensemble, $\ket{\psi_k}$ are the states of the ensemble on Bob's side, and $\lbrace \ket{k}_\text{A} \rbrace_{k=0}^{N-1}$ is the orthogonal basis of Alice's subspace.
After the transmission of the subsystem B, Alice conducts a von Neumann measurement associated with that basis, and this fixes a state of the ensemble on Bob's side.

For ensembles of mixed states, we have
\begin{equation}
    \rho_\text{AB} = \sum_{k=0}^{N-1} p_k \ket{k}\!\bra{k}_\text{A} \otimes \rho_k^\text{B},
\end{equation}
where $\rho_k^\text{B}$ are the density matrices of Bob's ensemble. 

Ideally, Alice and Bob share an entangled state of Eq.\,(\ref{eq:ent_state}).
During transmission through a quantum channel, quantum states can be modified by interactions with the environment, which may also generate entanglement between the system and the environment.
In quantum cryptographic protocols, the environment of a quantum channel is commonly assumed to be accessible to an adversary.
In the manner of the Lo-Chau protocol\,\cite{lo1999unconditional}, this scenario can be described in terms of the Stinespring’s theorem
\begin{equation}
    \ket{\Psi}_\text{ABE} = \sum_{k=0}^{N-1} \sqrt{p_k} \ket{k}_\text{A} \otimes \hat{U}\left( \ket{\psi_k}_\text{B} \otimes \ket{0}_\text{E} \right),
\end{equation}
where $\ket{0}_\text{E}$ is the initial state of the environment and $\hat{U}$ is the unitary operator that describes the interaction between transmitted state and the environment.

In the examples of the quantum operations below, we are especially interested in their ability to create ensemble coherence and how this ability is connected with the amount of information leaked to the environment.

\subsection*{Example 1. Measure-and-prepare transformation}\label{sec:intercet-resend}
Consider a transformation introduced in Ref.\,\cite{streltsov2011behavior} with the following Kraus operators
\begin{equation}
    K_0 = \ket{0}\!\bra{0}, K_1 = \ket{+}\!\bra{1}.
\end{equation}
It is straight forward to observe, this channel maps the input incoherent ensemble $\mathcal{E}_0 = \lbrace (\frac{1}{2}, \ket{0}), (\frac{1}{2}, \ket{1}) \rbrace$ of two orthogonal states to the following pair $\mathcal{E}_1 = \lbrace (\frac{1}{2}, \ket{0}), (\frac{1}{2}, \ket{+}) \rbrace$.
The resulting ensemble is a pair of non-orthogonal states, thus, coherent.

The channel considered here can generate ensemble coherence from a classical input ensemble.
Yet, conditioned on a particular Kraus outcome, every input with a nonzero probability of that outcome is mapped to the same predetermined state, yielding a classical ensemble.
This example illustrates why the definition of incoherent operations introduced in Section\,\ref{sec:basis} concerns the action of the full channel on the ensemble, rather than its individual Kraus branches.
The channel is also not an incoherent operation in the Baumgratz framework with respect to the computational basis.

The channel measures the input in the computational basis, then prepares $\ket{0}$ for outcome $0$ or $\ket{+}$ for outcome $1$.
To describe its action in the entanglement-based framework, suppose Alice prepares $\ket{\Psi}_\text{AB}=(\ket{00}+\ket{11})/\sqrt{2}$, keeps subsystem A, and sends subsystem B through the channel.
The resulting Alice--Bob state is
\begin{multline}
    (\mathds{1}_\text{A}\otimes\Phi_\text{B}) (\ket{\Psi}\!\bra{\Psi}) \\=\frac12\ket{0}\!\bra{0}_\text{A}\otimes\ket{0}\!\bra{0}_\text{B}
        +\frac12\ket{1}\!\bra{1}_\text{A}\otimes\ket{+}\!\bra{+}_\text{B}.
\end{multline}
Although this state is separable, Alice’s von Neumann measurement prepares the coherent ensemble $\mathcal E_1$ on Bob’s side.
The channel therefore generates ensemble coherence while destroying Alice--Bob entanglement.

The considered transformation can be associated with the intercept-resend attack.
In such a case, a malicious party can entangle their subsystem with the transmitted part of the initial entangled state and obtain the full information about encoded logical bits.
It is caused by the absence of the initial coherence in the ensemble $\mathcal{E}_0$.
Although Bob’s output ensemble has the maximum coherence attainable by two equiprobable pure qubit states\,\cite{kodukhov2025converting}, Eve knows the encoded bit perfectly.
Thus, even maximal coherence generated during transmission does not guarantee secrecy.
In the considered prepare-and-measure setting, nonzero coherence of the ensemble encoding the classical bits is necessary for secure key distribution.
This conclusion is formalized in the Section\,\ref{sec:coherence_bounds} below.

\subsection*{Example 2. Beam splitter-like transformation}
Here, we consider a beam splitter-like transformation that reduces the distinguishability of Bob’s signal states while transferring information about the encoded classical bits to the environment\,\cite{kronberg2025local}.
This operation is similar to beam splitting operation in quantum optics.

Consider an initial entangled state with two components,
\begin{equation}
    \ket{\Psi}_\text{AB} = \frac{1}{\sqrt{2}} \left( \ket{0}_\text{A} \otimes\ket{\psi_0}_\text{B} + \ket{1}_\text{A} \otimes \ket{\psi_1}_\text{B} \right),
    \label{eq:ensemble_B92}
\end{equation}
where $\ket{\psi_k} = \cos \alpha \ket{0}+(-1)^k \sin \alpha \ket{1} \,(k=0,1)$ are nonorthogonal states sent to Bob through the channel, $\braket{\psi_0 | \psi_1} = \cos 2\alpha$.
The transformation acts as
\begin{equation}
    \ket{\psi_k}_\text{B} \rightarrow \ket{\varphi_k}_\text{B} \otimes \ket{\varepsilon_k}_\text{E},
    \label{eq:BS}
\end{equation}
where the states $\ket{\varepsilon_0}$ and $\ket{\varepsilon_1}$ are leaked to the environment, which is controlled by a potential adversary, $\braket{\varepsilon_0|\varepsilon_1}=\cos 2\varepsilon$.
The states $\ket{\varphi_0}$ and $\ket{\varphi_1}$ form Bob's output ensemble, $\braket{\varphi_0 | \varphi_1} = \cos 2\beta$.
The Kraus operators acting on Bob's subsystem are
\begin{equation}
\begin{gathered}
    K_0 = \cos \varepsilon \left( \frac{\cos \beta}{\cos \alpha}\ket{0}\!\bra{0} + \frac{\sin \beta}{\sin \alpha} \ket{1}\!\bra{1}\right),\\
    K_1 = \sin \varepsilon \left( \frac{\sin \beta}{\cos \alpha}\ket{1}\!\bra{0} + \frac{\cos \beta}{\sin \alpha} \ket{0}\!\bra{1}\right).
\end{gathered}
\end{equation}
Preservation of inner products by the underlying unitary gives
\begin{equation}
    \braket{\psi_0 | \psi_1}
    = \braket{\varphi_0 | \varphi_1}\braket{\varepsilon_0 | \varepsilon_1}.
    \label{eq:unitary}
\end{equation}
It follows that Bob's output states are less distinguishable than the input states
\begin{equation}
    |\braket{\psi_0 | \psi_1}| \leq |\braket{\varphi_0 | \varphi_1}|.
\end{equation}
The channel is an incoherent operation in the Baumgratz framework with respect to the computational basis, since each Kraus operator maps diagonal states to diagonal states even upon postselection.
Nevertheless, as one can directly observe from numerical simulations below, for suitable parameters, it increases basis-free ensemble coherence and therefore is not an incoherent operation in the ensemble framework.

For the pure binary ensemble, low ensemble coherence occurs in two distinct limits: the input states nearly coincide, $|\braket{\psi_0 | \psi_1}|\rightarrow 1$, or they are nearly orthogonal, $|\braket{\psi_0 | \psi_1}|\rightarrow 0$.
In the first limit, little information about the encoded bit can be extracted.
In the second limit, the states can almost be distinguished by a von Neumann measurement.
The following example illustrates how this near-orthogonality allows substantial information leakage to the environment with little disturbance to the individual signal states.

\begin{figure}[t]
    \centering
    \includegraphics[width=\columnwidth]{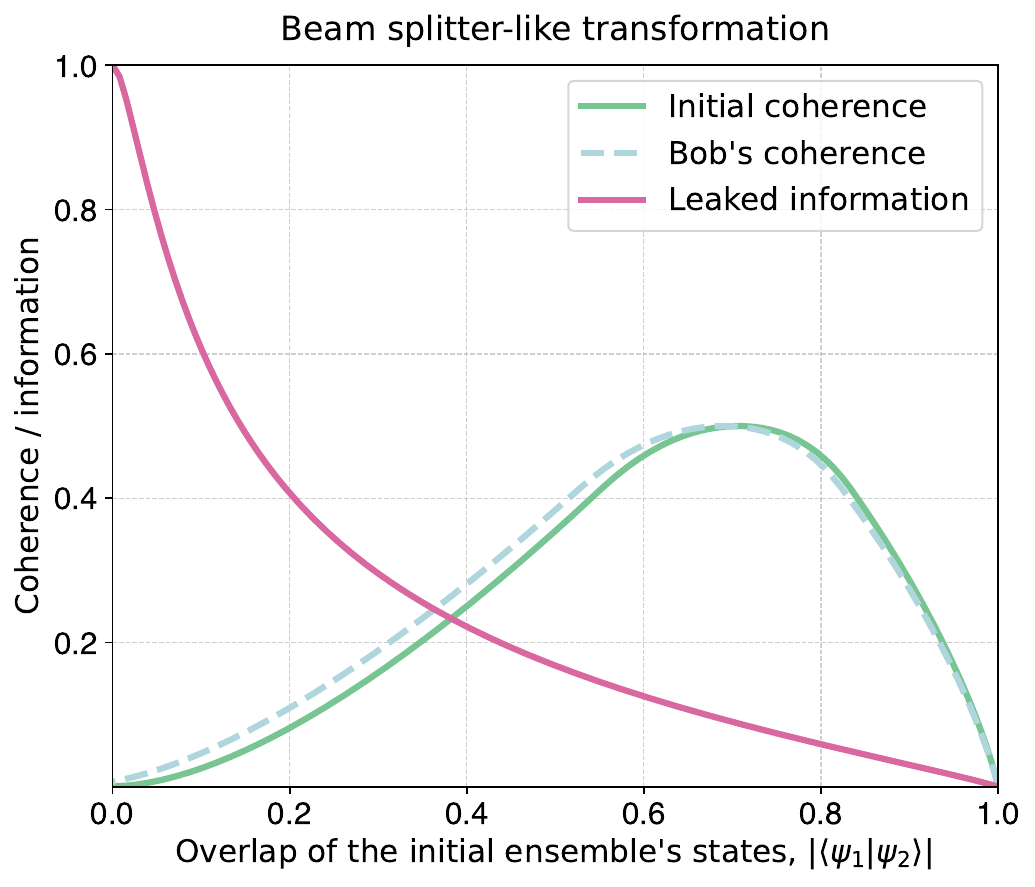}
    \caption{\textbf{Coherence under the beam splitter-like transformation.}
    Coherence and information leaked to the environment under a beam-splitter-like transformation of a pair of non-orthogonal states.
    The input and output ensemble coherences (green solid and teal blue dashed curves, respectively), together with the environment's Holevo quantity (pink curve), are plotted against the overlap between the states of the input ensemble.}
    \label{fig:beam-splitter-tradeoff}
\end{figure}

Figure\,\ref{fig:beam-splitter-tradeoff} shows the input and output ensemble coherences and the environmental information, determined by the Holevo quantity $\chi_\text{E}=h_2((1+\cos 2\varepsilon)/2)$, as functions of the overlap between the states of the input ensemble $|\braket{\psi_0 | \psi_1}|$.
For this illustration, we choose $\beta=\alpha/1.03$, corresponding to a small relative reduction of the angle between the ensemble states.
The channel therefore varies with the input angle $\alpha$, while the environmental overlap is determined by the unitarity condition in Eq.\,(\ref{eq:unitary}).
The environmental  Holevo quantity approaches one bit as the input states approach orthogonality.
For example, at $|\braket{\psi_0|\psi_1}|=0.01$, the input and output ensemble coherences are approximately $0.0004$ and $0.009$ bits, respectively, whereas $\chi_\text{E}\approx0.98$ bits.
Thus, when the input ensemble has small coherence, substantial information about the encoded classical bits can become available in the environment even though the individual signal states change only slightly.


The considered transformation models a beam splitter-like attack in a quantum cryptographic setting: Bob's states become less distinguishable while the environment acquires information about the classical variable encoded by Alice.
Opposing the Example\,1, the beam splitter-like transformation, given by Eq.\,(\ref{eq:BS}), does not fully eliminate the entanglement between Alice and Bob as long as output states are distinct.
Section\,\ref{sec:coherence_bounds} examines how postselection affects information extraction and secret key generation under this transformation.

The considered examples show that the operational value of ensemble coherence depends on the communication setting.
Coherence generated locally during Alice’s trusted preparation can provide a resource for subsequent information processing tasks.
However, coherence generated during transmission can coexist with information leakage to an adversary/environment, reduced signal distinguishability, or destruction of Alice–Bob entanglement.
Thus, a channel’s ability to increase ensemble coherence does not by itself establish an operational advantage: its significance must be assessed together with the preparation procedure, access to the environment, and the task being performed.
In the next section, we examine how ensemble coherence constrains information extraction with the postselective measurements and use the beam splitter-like transformation to derive an upper bound on the B92 QKD secret key rate.

\bigskip
\section{Coherence bounds on informational advantage and secret key generation}\label{sec:coherence_bounds}

In this section, we examine two operational roles of ensemble coherence: its relation to information extraction by postselective measurements, and its ability to bound the secret key rate of the qubit B92 QKD.
First, we show that coherent ensembles can be measured via the postselective measurement and the information conditioned on the conclusive measurement results can exceed the Holevo quantity.
Then, we show that the ensemble coherence for a pair of non-orthogonal states upper-bounds the effective informational advantage of unambiguous state discrimination over the Holevo quantity.
We then consider the beam splitter-like transformation of Example 2 for the B92 QKD protocol and derive an upper bound on the secret key rate in terms of Alice's ensemble coherence.

\subsection{Coherence and postselective measurements}
In the quantum communication scenarios, the essential characteristic of ensembles is the amount of extractable information.
The Holevo quantity bounds the information obtained from a quantum ensemble by means of nonselective collective measurements.
Postselection allows to overcome the Holevo bound for some measurement results, at the cost of some outcomes must be discarded.
Ref.\,\cite{kenbaev2022quantum} established sufficient conditions for such an informational advantage.
For a coherent ensemble of pure states, this conditional information advantage can be achieved by selecting a noncommuting pair and applying unambiguous state discrimination.

\textbf{Theorem (informational advantage).}
Let $\mathcal E=\{p_i,\ket{\psi_i}\}_{i=0}^{N-1}$ be an ensemble of pure states and $C(\mathcal E)>0$.
There exists a binary subensemble $\mathcal E_{km}$ and a postselective measurement such that the mutual information at conclusive measurement outcomes exceeds the Holevo quantity of the subensemble
\begin{equation}
    I_\text{USD}(\mathcal E_{km})>\chi(\mathcal E_{km}),
    \label{eq:ivdraft-existence}
\end{equation}
where the subscript “USD” denotes unambiguous state discrimination.
The Holevo quantity on the right refers to the selected subensemble before the measurement.

\textbf{Proof.}
Since ensemble has non-zero coherence, there exists at least one pair of distinct non-orthogonal states 
\begin{equation}
    \exists \ket{\psi_k}, \ket{\psi_m} \in \mathcal{E}:\,\,0 < |\!\braket{\psi_k | \psi_m}\!| < 1,
\end{equation}
where $k$ and $m$ are indexes of these states.
If there are several pairs of non-orthogonal states in the ensemble, the pair can be chosen randomly by a measurement party, Bob.
Bob applies the three-outcome POVM that unambiguously distinguishes between the states $\ket{\psi_k}$ and $\ket{\psi_m}$
\begin{equation}
    \begin{gathered}
        \Pi_k=\frac{\ket{\psi_m^\perp}\!\bra{\psi_m^\perp}}{1+|\!\braket{\psi_k | \psi_m}\!|},
        \Pi_m=\frac{\ket{\psi_k^\perp}\!\bra{\psi_k^\perp}}{1+|\!\braket{\psi_k | \psi_m}\!|},\\
        \Pi_{?}=\mathds{1}-\Pi_k-\Pi_m,
    \end{gathered}
    \label{eq:POVM-usd}
\end{equation}
where the states $\ket{\psi_{k(m)}^\perp}$ lie in $\operatorname{span}\{\ket{\psi_k},\ket{\psi_m}\}$ and are orthogonal to $\ket{\psi_{k(m)}}$.

Bob announces the chosen subensemble and whether he obtained a conclusive measurement result.
Alice, who encodes classical information in the initial ensemble, announces whether the prepared state belongs to the selected subensemble.
Alice and Bob keep only those rounds in which the prepared state belongs to the chosen pair and Bob obtains a conclusive outcome.

When the pair is fixed, the probabilities inside the subensemble $\mathcal{E}_{km}$ are $q=p_k/(p_k+p_m)$ and $1-q$.
Both states have the same success probability, $p_\checkmark = 1-|\!\braket{\psi_k | \psi_m}\!|$, so conclusive outcome does not change these probabilities.
The conclusive outcomes identify the classical encoded bit without error, giving
\begin{equation}
    I_\text{USD}=h_2(q).
\end{equation}
For this pure-state pair,
\begin{multline}
    \chi(\mathcal E_{km})=
    h_2\!\left(\frac{1+\sqrt{1-4q(1-q)(1-|\!\braket{\psi_k | \psi_m}\!|^2)}}{2}\right)\\
    <h_2(q).
    \label{eq:ivdraft-pair-holevo}
\end{multline}
The strict inequality follows from $0<q<1$ and $0<|\!\braket{\psi_k | \psi_m}\!|<1$.
The fraction of rounds where Bob successfully identifies the classical encoded bits is $(p_k+p_m)p_\checkmark$.
\qed

There is no violation of the Holevo bound: including the inconclusive outcomes gives    $p_\checkmark h_2(q) < \chi(\mathcal E_{km})$.
As a result, the theorem claims that, for a coherent ensemble, it is possible to choose a subensemble of non-commuting states.
For such subensemble, one can conduct a postselective measurement and achieve informational advantage over the Holevo quantity of the subensemble. 

\subsection{Coherence bounds the effective informational advantage}

We now quantify the postselected informational advantage over the Holevo quantity.
For two equiprobable pure states, ensemble coherence bounds this informational advantage multiplied by the probability of a conclusive outcome.

\textbf{Proposition 1 (Effective informational advantage and coherence).}
Let $\mathcal{E}=\{(1/2,\ket{\psi_i})\}_{i=0,1}$ consist of two distinct non-orthogonal pure states of finite dimension $d$.
For the symmetric optimal USD, Eq.\,(\ref{eq:POVM-usd}), the ensemble coherence upper-bounds the effective informational advantage
\begin{equation}
    p_\checkmark\bigl(I_\text{USD}(\mathcal{E})-\chi(\mathcal{E})\bigr)<C(\mathcal{E}),
    \label{eq:inf_adv}
\end{equation}
where $p_\checkmark=1-|\!\braket{\psi_0|\psi_1}\!|$ is the probability of a conclusive outcome and $I_\text{USD}=1$ bit is the information per conclusive outcome.

\textbf{Proof.}
Write $\rho_i=\ket{\psi_i}\!\bra{\psi_i}$ and fix an arbitrary orthonormal basis $T=\{\ket{e_k}\}_{k=0}^{d-1}$.
Let $\rho_i^\text{diag}$ be the diagonal part of $\rho_i$ in this basis, and let $X_i=\rho_i-\rho_i^\text{diag}$ be its off-diagonal part.

For any probability distribution $\{p_k\}$, the Harremo\"es--Tops\o e inequality\,\cite{harremoes2001inequalities} gives
\begin{equation}
    H(\{p_k\})\geq2\left(1-\sum_kp_k^2\right),
    \label{eq:entropy_collision_lower_bound}
\end{equation}
see Eq.\,(11) in Ref.\,\cite{wu2009entropic}.
Additionally, we need the norm of the operators $X_i$
\begin{equation}
    \|X_i\|^2 = \text{tr}(X_i^\dagger X_i) = 1-\sum_{k=0}^{d-1}|\!\braket{e_k|\psi_i}\!|^4, \,i=0,1.
\end{equation}
Applying Eq.\,(\ref{eq:entropy_collision_lower_bound}) to the expression of the ensemble coherence in the basis $T$ gives
\begin{equation}
        \!\!C_T(\mathcal{E})
        =\frac12\sum_{i=0,1} H\!\left(\{|\!\braket{e_k|\psi_i}\!|^2\}_k\right)
        \!\geq\!\|X_0\|^2\!+\!\|X_1\|^2.
    \label{eq:coherence_offdiagonal_lower_bound}
\end{equation}
Since $\rho_0^\text{diag}$ and $\rho_1^\text{diag}$ are diagonal in the same basis, they commute, and hence
\begin{equation}
    [\rho_0,\rho_1]=[X_0,\rho_1]+[\rho_0^\text{diag},X_1].
\end{equation}
For any density matrix $\rho$ and any operator $A$, one has $\|[A,\rho]\|\leq\|A\|$.

Both $\rho_1$ and $\rho_0^\text{diag}$ are density operators, so the triangle inequality gives
\begin{equation}
    \|[\rho_0,\rho_1]\|\leq\|X_0\|+\|X_1\|.
\end{equation}
The norm of the commutator of two density matrices is no greater than the sum of the norms of their off-diagonal parts.
Squaring, we obtain
\begin{equation}
    \|[\rho_0,\rho_1]\|^2
    \leq2\bigl(\|X_0\|^2+\|X_1\|^2\bigr).
\end{equation}
For the two pure states, direct calculation gives
\begin{equation}
    \|[\rho_0,\rho_1]\|^2=2s^2(1-s^2),
\end{equation}
where $s=|\!\braket{\psi_0|\psi_1}\!|$.
Combining these estimates with Eq.\,(\ref{eq:coherence_offdiagonal_lower_bound}) and minimizing over all bases proves the following lower-bound
\begin{equation}
    C(\mathcal{E})\geq s^2(1-s^2).
    \label{eq:pure_pair_coherence_lower_bound}
\end{equation}

The Holevo quantity of the equiprobable pure-state pair is $\chi(\mathcal{E})=h_2((1+s)/2)$.
Then, we exploit the bound on the binary entropy\,\cite{topsoe2001bounds}
\begin{equation}
    h_2\left(\frac{1+x}{2}\right)\geq1-x^2.
    \label{eq:h2_lower_bound}
\end{equation}
We apply it to the effective informational advantage, i.e. the left-hand side of Eq.\,(\ref{eq:inf_adv}), and obtain
\begin{multline}
    p_\checkmark\bigl(I_\text{USD}(\mathcal{E})-\chi(\mathcal{E})\bigr)
        =(1-s)\left(1-h_2\left(\frac{1+s}{2}\right)\right)\\
        \leq(1-s)s^2 < (1-s^2)s^2\leq C(\mathcal{E}),
\end{multline}
where the strict inequality follows from $s<1$.
This completes the proof.
\qed

One can directly study the asymptotic of Eq.\,(\ref{eq:inf_adv}).
If the input ensemble tends to a pair of orthogonal states, i.e. to an ensemble with zero coherence, the Holevo quantity tends to 1, approaching the $I_\text{USD}$, and the informational advantage disappears.
If the input states almost coincide, $s \rightarrow 1$, the probability of a conclusive measurement results vanishes despite the informational advantage is non-zero.
Ensemble coherence thus provides a quantitative resource constraint on postselected information extraction: as coherence vanishes, either the conditional informational advantage or the probability of success must also vanish.
Importantly, the Eq.\,(\ref{eq:inf_adv}) works for any finite dimension.

\subsection{Ensemble coherence and the key rate of the B92 QKD}
We now use the beam splitter-like transformation of Example 2 to connect initial ensemble coherence with a key rate for the B92 QKD protocol.

We consider the B92 QKD protocol in the following form: 
Alice encodes equiprobable classical bits $k=0,1$ into a pure qubit ensemble $\mathcal{E}_\text{B92}$ defined in Eq.\,(\ref{eq:ensemble_B92}).
Then, Alice transmits prepared states through the quantum channel.
Bob performs symmetric optimal USD for the arrived states and discards all inconclusive measurement outcomes.
Bob announces for which transmitted states he obtained conclusive outcomes over an authenticated public channel, while keeping the decoded bits private.
Then, legitimate users conduct classical post-processing including error correction and privacy amplification.

As a quantum channel, we consider the beam splitter-like transformation that makes the states of the initial ensemble less distinguishable and leaks quantum states to the environment as expressed in Eq.\,(\ref{eq:BS})
\begin{equation}
    \ket{\psi_k}_\text{B} \rightarrow \ket{\varphi_k}_\text{B} \otimes \ket{\varepsilon_k}_\text{E}.
    \label{eq:BS_S4}
\end{equation}
Eve stores all environmental states and may measure them collectively after the public communication.
The scalar products of the input pair, output pair and the environment states must satisfy
\begin{equation}
         \braket{\psi_0 | \psi_1} = \braket{\varphi_0 | \varphi_1}\braket{\varepsilon_0 | \varepsilon_1} .
\end{equation}
The latter condition on the scalar products is the requirement of unitary transformation over Bob-Eve subsystems.
The corresponding Kraus representation is given in Example 2.
Under the considered attack, Bob’s raw bits are error-free when the devices are ideal.
The probability of a conclusive outcome is $p_\checkmark=1-|\braket{\varphi_0|\varphi_1}|$, regardless of which bit Alice sends.
Because the output states in Eq.\,(\ref{eq:BS_S4}) are products, announcing success in USD neither changes Eve's conditional states nor biases the probability distribution of logical bits in the raw data.

Alice-Bob-Eve classical-quantum state after the postselection is described as
\begin{equation}
    \rho_\text{ABE$|\checkmark$}
    =\frac12\sum_{k=0,1}
    \ket{kk}\!\bra{kk}_\text{AB}\otimes
    \ket{\varepsilon_k}\!\bra{\varepsilon_k}_\text{E}.
    \label{eq:ABE-state}
\end{equation}
Eve's information is upper-bounded by the Holevo quantity
\begin{equation}
    I(\text{A,E}) \leq \chi_\text{E}=h_2\!\left(\frac{1+|\!\braket{\varepsilon_0 | \varepsilon_1}\!|}{2}\right).
    \label{eq:Eve_BS}
\end{equation}
According to the Devetak-Winter theorem\,\cite{devetak2005distillation}, the resulting asymptotic secret key rate is
\begin{equation}
    K_\text{BS} = p_\checkmark \left(1-h_2\!\left(\frac{1+|\!\braket{\varepsilon_0 | \varepsilon_1}\!|}{2}\right)\right).
    \label{eq:BS-key-rate}
\end{equation}

In the described protocol, we assume that Eve has no access to the devices of legitimate users and the devices do not leak any side-channel information, while the quantum channel is under full Eve's control.
Let $K_\text{B92}$ denote the secure key rate of the described B92 protocol against all attacks consistent with the assumptions stated above.
Since the beam splitter-like attack is one particular attack allowed under these assumptions, the secure key rate of the considered B92 protocol cannot exceed the rate $K_\text{BS}$ under this attack
\begin{equation}
    K_\text{B92} \leq K_\text{BS}.
    \label{eq:rate-UB}
\end{equation}
We now bound this secret key rate by the coherence of Alice's initial ensemble.

\textbf{Proposition 2 (Coherence and B92 QKD key rate).}
For the B92 QKD and attack defined above, suppose the scalar product between the output states satisfies the following lower bound
\begin{equation}
    |\!\braket{\varphi_0|\varphi_1}\!|\geq \frac{\sqrt5-1}{2}.
    \label{eq:output-condition}
\end{equation}
Then
\begin{equation}
    K_\text{BS} \leq C(\mathcal{E}_\text{B92}).
    \label{eq:ivdraft-initial-coherence-rate}
\end{equation}
The same upper bound holds for $K_\text{B92}$.

\textbf{Proof.}
In the computational basis, the input states can be written as
\begin{equation}
    \ket{\psi_k}=\cos\alpha\ket{0}+(-1)^k\sin\alpha\ket{1},
\end{equation}
where $s=|\!\braket{\psi_0|\psi_1}\!|=\cos(2\alpha)$ and $0<\alpha<\pi/4$.
Consider an arbitrary orthonormal basis $T=\{\ket{e_0}\!,\ket{e_1}\}$, where
\begin{equation}
    \ket{e_0}=\cos\theta\ket{0}+e^{i\phi}\sin\theta\ket{1},
\end{equation}
with $0\leq\theta\leq\pi/2$ and $0\leq\phi<2\pi$, and $\braket{e_0|e_1}=0$.
The probabilities $p_k=|\!\braket{e_0|\psi_k}\!|^2$ satisfy
\begin{equation}
    2p_k-1=s\cos(2\theta)
    +(-1)^k\sqrt{1-s^2}\sin(2\theta)\cos\phi.
    \label{eq:qubit-basis-probabilities}
\end{equation}
Using Eq.\,(\ref{eq:h2_lower_bound}) for each contribution to the ensemble coherence and removing the terms with opposite signs, yielding
\begin{multline}
    C_T(\mathcal{E}_\text{B92})
    =\frac12\left( h_2(p_0)+h_2(p_1)\right)\\
    \geq1-s^2\cos^2(2\theta)-(1-s^2)\sin^2(2\theta)\cos^2\phi\\
    \geq 1-\max\{s^2,1-s^2\}.
    \label{eq:qubit-basis-coherence-bound}
\end{multline}
The last inequality uses $\cos^2\phi\leq1$ and $\cos^2(2\theta)+\sin^2(2\theta)=1$.
Since this bound holds for every basis $T$, minimization gives
\begin{equation}
    C(\mathcal{E}_\text{B92})\geq\min\{s^2,1-s^2\}.
    \label{eq:ivdraft-qubit-lower-bound}
\end{equation}
Applying Eq.\,(\ref{eq:h2_lower_bound}) to Eve's information, Eq.\,(\ref{eq:Eve_BS}), we obtain the uperbound on the key rate in the case of the beam splitting-like attack
\begin{equation}
    K_\text{BS} \leq (1-|\!\braket{\varphi_0|\varphi_1}\!|)\frac{s^2}{|\!\braket{\varphi_0|\varphi_1}\!|^2}\leq s^2,
\end{equation}
where the last inequality follows from the proposition condition, Eq.\,(\ref{eq:output-condition}).
Independently, one can show
\begin{equation}
    K_\text{BS} \leq 1-|\!\braket{\varphi_0|\varphi_1}\!| \leq1-s\leq1-s^2.
\end{equation}
Combining both estimates with Eq.\,(\ref{eq:ivdraft-qubit-lower-bound}) proves
\begin{equation}
    K_\text{B92} \leq K_\text{BS}
    \leq\min\{s^2,1-s^2\}\leq C(\mathcal{E}_\text{B92}).
    \label{eq:ivdraft-security-coherence-bound}
\end{equation}
\qed

We have shown that, for the parameter range specified in Proposition 2, the coherence of Alice’s initial ensemble upper-bounds the secret key rate under the beam splitter-like attack.
For the channel conditions considered here, the guaranteed secure key rate $K_\text{B92}$ is therefore also upper-bounded by the coherence of Alice’s initial ensemble.

Proposition 2 applies, for example, to the beam splitter-like transformation of Example 2 with $\beta=\alpha/1.75$.
For this choice, the output overlap $\cos(2\beta)$ satisfies Eq.\,(\ref{eq:output-condition}) for every $0<\alpha<\pi/4$, so the initial ensemble coherence upper-bounds the secret key rate throughout this family.

The two propositions established in this section have a similar mathematical form but different operational meanings.
Proposition 1 bounds the effective informational advantage of postselection over the Holevo quantity for a pair of pure states in any finite dimension.
Proposition 2 bounds the secret key rate of the qubit B92 QKD protocol under the stated assumptions.
Together, they give ensemble coherence a quantitative operational interpretation: it limits both the effective gain in information obtained through postselection and, in the cryptographic setting considered here, the rate at which secret key can be generated.

\bigskip
\section{Conclusion}\label{sec:conclusion}
In this work, we have extended the study of basis-free ensemble coherence in two complementary directions: its resource-theoretic properties and its operational role in quantum communication.

Building on Haar-averaged coherence\,\cite{singh2016average,cheng2015complementarity,zhang2017average}, we established a tight upper bound on relative entropy ensemble coherence that depends only on the Hilbert space dimension.
The continuous ensemble of all pure states distributed according to the normalized Haar measure attains this bound and is therefore maximally coherent in every finite dimension.
This bound can be tighter than the Holevo quantity and places a limit on the conversion of entanglement into ensemble coherence studied in Refs.\,\cite{kodukhov2023entanglement,kodukhov2025converting}.
In particular, since the maximal ensemble coherence in dimension $d$ is smaller than $\log_2 d$, the full entanglement of a maximally entangled state cannot be converted into ensemble coherence.

We identified commutativity-preserving channels\,\cite{yu2011quantum,streltsov2011behavior,hu2012necessary} as the nonselective incoherent operations for ensembles, with the required monotonicity of relative entropy ensemble coherence established previously in Ref.\,\cite{mao2019quantumness}.
This definition differs from incoherent operations in the Baumgratz framework\,\cite{baumgratz2014quantifying}, which admit a Kraus representation in which every branch preserves states diagonal in a fixed reference basis.
For ensemble coherence, the defining requirement is that the full channel map every ensemble of mutually commuting states to another ensemble of mutually commuting states, without fixing a reference basis or imposing this requirement on individual Kraus branches.
The beam splitter-like transformation of Example 2 illustrates this distinction: it is incoherent in the Baumgratz framework but can generate basis-free ensemble coherence.

Building on the study of postselective informational advantage\,\cite{kenbaev2022quantum}, we established a quantitative connection between ensemble coherence and information extraction by postselective measurements.
Every coherent ensemble of pure states contains a pair for which unambiguous discrimination yields more information per conclusive outcome than the Holevo quantity of the selected pair.
Proposition 1 bounds this conditional informational advantage, weighted by its success probability, by the coherence of an equiprobable pair of non-orthognal states in any finite dimension.
Thus, as ensemble coherence tends to zero, either the informational advantage per conclusive outcome or the probability of obtaining such an outcome must tend to zero.

Taking this operational perspective into quantum cryptography, we derived a coherence-based upper bound on the asymptotic secret key rate of the considered qubit B92 QKD protocol\,\cite{bennett1992quantum}.
Proposition 2 bounds the key rate under the beam splitter-like attack by the coherence of Alice’s initial ensemble, provided that the overlap between Bob’s output states satisfies the lower bound in Eq.\,(\ref{eq:output-condition}).
Since the guaranteed secure rate of the considered protocol cannot exceed its rate under a particular attack, the same coherence bound applies to the actual key generation rate.
Ensemble coherence therefore acquires a direct quantitative cryptographic meaning as a limit on secret key generation.

The cryptographic result established here concerns the qubit B92 QKD protocol, leaving open the broader role of ensemble coherence in QKD.
An important direction is to determine the conditions under which nonzero ensemble coherence is necessary for secrecy and what additional assumptions would make it sufficient to enable secret key generation.
The coherence entering Proposition 2 depends only on Alice's encoding ensemble, whereas a complete protocol also specifies Bob's measurement and the classical postprocessing.
Clarifying how these choices interact with the initial coherence is essential for extending the present connection to wider classes of quantum communication protocols.

\bibliography{re}
\end{document}